\documentclass[runningheads]{llncs}

\usepackage{graphicx}
\usepackage{subfig}
\usepackage{dsfont}
\usepackage{color,amsfonts,amsmath,algorithm,algpseudocode,tikz,verbatim}
\usepackage{enumerate}% http://ctan.org/pkg/enumerate
\usepackage[percent]{overpic}
\usepackage{xcolor}
\usepackage{hhline}
\usepackage{array}
\newcolumntype{C}[1]{>{\centering\let\newline\\\arraybackslash\hspace{0pt}}m{#1}}

\begin{document}

\title{A Novel Data-driven Algorithm for the Automated Detection of Unexpectedly High Traffic Flow in Uncongested Traffic States
}

\author{Bo Klaasse\inst{1} \and
Rik Timmerman\inst{1}\orcidID{0000-0002-8964-8748} 
\and
Tessel van Ballegooijen\inst{2} \and
Marko Boon\inst{1}\orcidID{0000-0001-8804-8743} \and
Gerard Eijkelenboom\inst{2}}

\authorrunning{B. Klaasse et al.}
\titlerunning{Unexpectedly High Traffic Flow in Uncongested Traffic States}% Part of RIGHT running header

\institute{Eindhoven University of Technology, Eindhoven, The Netherlands \email{r.w.timmerman@tue.nl} 
\and 
De Verkeersonderneming, Rotterdam, The Netherlands}
\maketitle              % typeset the header of the contribution
\begin{abstract}
We present an algorithm to identify days that exhibit the seemingly paradoxical behaviour of high traffic flow and, simultaneously, a striking absence of traffic jams. We introduce the notion of high-performance days to refer to these days. The developed algorithm consists of three steps: step 1, based on the fundamental diagram (i.e. an empirical relation between the traffic flow and traffic density), we estimate the critical speed by using robust regression as a tool for labelling congested and uncongested data points; step 2, based on this labelling of the data, the breakdown probability can be estimated (i.e. the probability that the average speed drops below the critical speed); step 3, we identify unperturbed moments (i.e. moments when a breakdown is expected, but does not occur) and subsequently identify the high-performance days based on the number of unperturbed moments.
Identifying high-performance days could be a building block in the quest for traffic jam reduction; using more detailed data one might be able to identify specific characteristics of high-performance days.
The algorithm is applied to a case study featuring the highly congested A15 motorway in the Netherlands. 

\end{abstract}

\section{Introduction} \label{sec:intro}

Nowadays, traffic jams have become an inevitable part of road traffic. In particular, near or in urban areas the high vehicle-to-capacity ratio on the road imposes cars to slow down or even stop too frequently. This causes a wide variety of problems, as in the Netherlands alone, the amount of monetary value lost due to traffic jams in 2018 is estimated at 1.3 billion euros~\cite{AD2019}. Moreover, traffic jams cause pollution and decrease the quality of life e.g. in cities.

Reducing traffic congestion is a challenging problem. Obviously, increasing the capacity of existing roads, e.g. by adding lanes, would provide a solution to the problem. However, such actions are costly and not always desired, or even possible. Alternatively, one could aim to influence drivers' behaviour. This can be achieved by, for example, monetary means (such as toll systems or congestion pricing, see e.g.~\cite{bergendorff1997congestion,goh2002congestion}), encouraging drivers to drive outside peak hours (see~\cite{ettema2010using} for instance) or dynamic road signalling (see e.g.~\cite{hegyi2008specialist}). It is increasingly important to find the exact effect of these measures, but this is a complicated problem, which is partly due to the highly complex nature of traffic and the fact that the manifestation of congestion is subject to randomness, see for example~\cite{arnesen2017estimator,tu2008monitoring}.

In this paper, we approach the problem of reducing traffic congestion from a different perspective, as we look at the \emph{absence} of traffic jams. Typically, once the traffic flow, i.e. the throughput measured in vehicles per hour, has passed a certain threshold, congestion \emph{could} emerge. This phenomenon is referred to as a ``breakdown''. We are interested in days during which a relatively large number of breakdowns were expected, but did not occur. Such days will be referred to as ``high-performance days''. Specifically, we develop an algorithm to automatically identify these high-performance days based on historical traffic data and test our method on a section of the A15 motorway in the Netherlands. In a future study, one could try to determine the specific characteristics of the resulting high-performance days using more detailed data. Ultimately, the goal is to find out whether the high-performance days could be caused by specific  behavioural patterns of individual drivers. However, we focus on the first step, namely the automated detection of high-performance days.

Our algorithm relies on the shape of the fundamental diagram, the well-known empirical diagram that displays the relationship between the traffic flow $q$ (vehicles per hour) and the traffic density $\rho$ (vehicles per kilometre) at a specific location. Many studies have shown that the fundamental diagram can be divided into two regions, a region for congestion and a region for free flow. The empirical fundamental diagram has been studied extensively and a wide variety of theoretical models has been proposed (see for example~\cite{Gaddam2019} for an overview). %Whereas the flow-density relationship during free flow can be modeled by a curve with a strictly positive slope, this is not the case during congestion. Therefore, most models differ in how they handle the congestion region.  
Our aim is not a theoretical model for the fundamental diagram however; we are merely interested in the critical speed, i.e. the speed which defines congestion and separates the free-flow region from the congestion region in the fundamental diagram. So, we can get around the problem of modelling the congestion region and exploit the roughly linear flow-density relationship during free flow. We show that \emph{robust regression} is an excellent technique to obtain the free-flow speed and subsequently distinguish between free flow and congestion based on the calculated weights. Utilizing the method proposed by~\cite{arnesen2017estimator}, we subsequently estimate the \textit{breakdown probability}. This paves the way to identifying high-performance days.

To the best of our knowledge, our approach to obtain the critical speed and the introduction of the notion of high-performance days are original. Many papers focus on (real-time) traffic jam estimation using GPS-data and/or trajectory data, see a.o.~\cite{ong2011traffic,petrovska2015traffic,vaqar2009traffic}. This is partly due to the widespread availability of GPS data. However, we have chosen to use detector data, as traffic detectors are present on most Dutch motorways and provide a sufficiently high granularity. Detector data is also used in the literature; in~\cite{li2010comparison} detector data is used to automatically track congestion and in~\cite{kerner1997experimental} detector data is used to study phase transitions on German highways. However, the work that is probably closest to our study is~\cite{dervisoglu2009automatic}. Therein, the authors use detector data to estimate highway characteristics such as the free-flow speed and the critical density. These quantities are then used to calibrate a cell transmission model. We determine a related highway characteristic (the critical speed), but in our study this is a tool to estimate the breakdown probability. Indeed, our main goal is different: we identify a surprising \emph{absence} of traffic jams. This could be an important first step towards a better understanding of the reasons why on certain days the traffic flow is so much better than on other days, although the circumstances seem to be identical.

%\subsubsection{Organization of the Paper.}

In Section~\ref{sec:setting} we provide information about the location of the experimental region and discuss the data. We proceed with the theoretical foundation and the three main steps of the algorithm in Section~\ref{sec:theory}. The validation of important assumptions and parameter choices is presented in Section~\ref{sec:results}, as well as the main insights of the case study. We close with a conclusion and multiple suggestions for future research in Section~\ref{sec:concl}.

\section{Description of the Location and the Data} \label{sec:setting}

In this section, we discuss the relevant aspects of the part of the A15 motorway from which the data is obtained. Subsequently, we elaborate on the structure of the data set and which steps we take in the preprocessing of the data.

\subsection{Location of the Experimental Region}\label{sec:location}
\begin{figure}[ht]%
 \centering
 \subfloat[]{
    \begin{overpic}[width=0.43\linewidth]{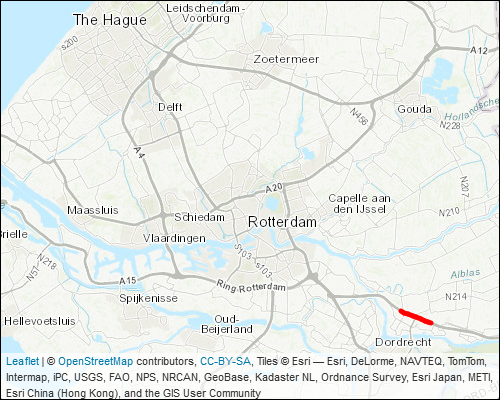}
        \newcommand{\bignwarrowred}{\textcolor{red}{\text{\scalebox{2}{\ensuremath{\downarrow}}}}}
        \put(80,22){$\bignwarrowred$}
    \end{overpic}\label{fig:1a}}%
 \quad
 \subfloat[]{
    \begin{overpic}[width=0.43\linewidth]{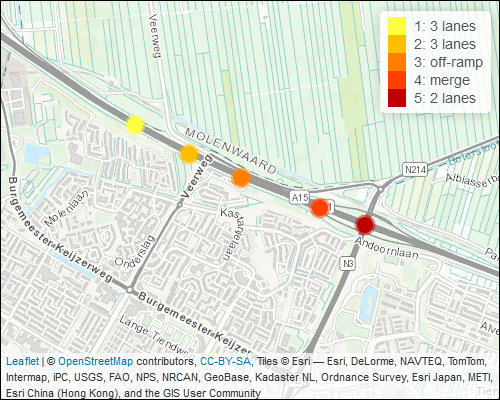}
    \end{overpic}\label{fig:1b}}\\%
 \caption{(a) Overview of the trajectory, marked red and indicated by the red arrow, in relation to Rotterdam. (b) The location of the five detectors on the trajectory.}%
 \label{fig:location}%
\end{figure}

The location under consideration is the A15 motorway near Rotterdam, at the interchange with Papendrecht (see Figure~\ref{fig:location}). Five detectors have been placed in the eastern direction, with a distance of approximately 300 metres between consecutive detectors (see Figure~\ref{fig:location}b). Between the second and third detector, an off-ramp to Papendrecht is located. Shortly afterwards, the vehicles on the A15 merge from three to two lanes. 
%In Figure~\ref{fig:schematic} a schematic overview of the experimental region is presented. 
The maximum speed along this whole trajectory of the A15 is 120 km/h. The traffic jams on this trajectory belong to the most costly traffic jams in the Netherlands (see \cite{AD2019}) and the A15 is one of the most congested roads in the Netherlands, connecting one of the world's largest ports with the European main land, which makes this a particularly interesting motorway to study.

%\begin{figure}[ht!]
%    \centering
%    \includegraphics[width =0.8 \linewidth]{schematic_location_crop.pdf}
%    \caption{A schematic overview of the experimental region.}
%    \label{fig:schematic}
%\end{figure}

\subsection{Description of the Data Set}\label{sec:data}

The data is obtained from the Dutch National Data Warehouse for Traffic Information (NDW), a collaboration of 19 public authorities that cooperate on collecting, storing, and redistributing data. The data is publicly available and can be requested at the website of the NDW~\cite{NDW}. The data we obtained from the NDW spans a period from January 1, 2018 until December 31, 2018. Every minute, the detectors measure, for each lane individually, the number of vehicles that have passed (i.e the traffic flow $q$, in vehicles per hour) and the average speed $v$ of the passing cars in kilometres per hour, calculated using the arithmetic mean. We can estimate the average traffic density $\rho$ using $\rho=q/v$, although this formula is known to underestimate the density when the arithmetic mean is used~\cite{knoop2017automatic}. We combine the various lanes as in~\cite{treiber2013traffic}. For the sake of reducing the variability in the data, we aggregate the measurements to a period of 5 minutes, as is done in~\cite{arnesen2017estimator}. The arithmetic mean is used to obtain the average traffic flow and the average speed is calculated analogous to the average speed over multiple lanes. 

The resulting data set can be described as follows. We introduce the set of locations $\mathcal{I}:=\lbrace 1,2,3,4,5\rbrace$, in accordance with Figure~\ref{fig:location}b.
%We do not use the data from location 6 and location 7 as those detectors were (partially) malfunctioning throughout the first half of the year. Also, we do not use data from locations where the speed limit is not constant, so we also exclude locations 8 and 9. 
%Moreover, due to malfunctioning detectors, some measurements are missing, or incomplete. Because our analysis relies on the shape of the fundamental diagram (see Section~\ref{sec:critical_speed}), we decided to only include days with at least 20 hours of full data. 
Moreover, we focused our research on weekdays and thereby excluded all weekend days from the data, because the traffic flow is oftentimes significantly lower. The set containing all 261 weekdays in 2018 is denoted by $\mathcal{J}$. After the aforementioned exclusions, we have one set of measurement dates $\mathcal{J}^{(i)} \subseteq \mathcal{J}$ for each detector $i \in \mathcal{I}$. At each location we have measurements of the average traffic flow and average vehicle speed, as well as an estimate for the density, aggregated to 5-minute intervals. Hence, for location $i\in \mathcal{I}$ and date $j \in \mathcal{J}^{(i)}$ we have a sequence of measurement times 
\begin{equation}
    \mathcal{T}^{(i,j)}:= \Big\{  t_1^{(i,j)}, t_2^{(i,j)}, \ldots\Big\}\subseteq \mathcal{T},
\end{equation}
where $\mathcal{T}$ is the set containing all 5-minute intervals on a day. The corresponding set of measurements for detector $i$ on date $j$ is
\begin{equation}
    \mathcal{X}^{(i,j)}:=\bigg\{ \left(q^{(i,j)}_t, v^{(i,j)}_t, \rho^{(i,j)}_t \right) : t \in \mathcal{T}^{(i,j)}\bigg\}.
\end{equation}
The data set containing only the flow and the density is denoted by
\begin{equation}
    \bar{\mathcal{X}}^{(i,j)}:=\bigg\{ \left(\rho^{(i,j)}_t, q^{(i,j)}_t \right) : t \in \mathcal{T}^{(i,j)}\bigg\}.
\end{equation}

In total we have $\vert\mathcal{I}\vert=5$ locations and $\vert\mathcal{J}\vert=261$ dates, leading to a total of $5\cdot 261=1305$ instances. However, in the first step of the algorithm (i.e. estimating the critical speed), we do not include all days/critical speeds:
\begin{enumerate}
    \item We exclude the most extreme critical speeds of each location (see Section~\ref{sec:theory} for a motivation in relation to our assumptions and Equation~\eqref{eq:2sigma}/the last paragraph of Section~\ref{sec:robust} for a further elaboration);
    \item We exclude instances where the free-flow and congestion region are not linearly separable by a straight line through the origin, given the labelling (see Remark~\ref{rem:not_separable});
    \item We exclude days with little or no congestion (see Section~\ref{sec:validation}).
\end{enumerate}
For the remaining steps, we do include all 1305 instances, meaning that no weekdays are beforehand excluded when identifying the high-performance days.

All the analyses were performed in the statistical software package~R.

\section{The Main Algorithm} \label{sec:theory}
We present the main algorithm in this section and elaborate on the theoretical foundation using traffic theory, robust regression and the estimator for the breakdown probability proposed in~\cite{arnesen2017estimator}. The algorithm consists of three parts: (i) estimating the critical speed, (ii) estimating the breakdown probability, and (iii) identifying the high-performance days. In Section~\ref{sec:critical_speed} we formally define the relevant notions, such as the critical speed. In Section~\ref{sec:robust} we explain how the critical speed is obtained using robust regression as a labelling tool. Lastly, in Section~\ref{sec:breakdown_high-performance} we discuss the estimator for the breakdown probability and provide a definition for high-performance days, based on ``unperturbed moments''.

\subsection{The Fundamental Diagram and the Critical Speed} \label{sec:critical_speed}
Studying the traffic behaviour at a specific location, say location $i$, one can distinguish two different traffic states: free flow and congestion. As in~\cite{kerner2009IntroModern}, we can define free flow and congestion based on the critical speed. 
\begin{definition}[Free flow, Congestion and Critical speed]\label{def:free_flow_and_congestion}
\emph{Free (traffic) flow} is a state when the vehicle density
in traffic is small enough for interactions between
vehicles to become negligible. Therefore, vehicles have an opportunity to move at their desired maximum speeds \cite{kerner2009IntroModern}. 
When the density increases beyond a certain threshold in free flow, vehicle interaction cannot be neglected anymore. Due to this vehicle interaction, the average vehicle speed decreases to a value lower than the \emph{critical speed}, which is the minimum average speed that is still possible in free flow. This new state of traffic is referred to as a state of \emph{congested traffic}. 
\end{definition}

We denote the critical speed at location $i$ by $v^{(i)}_{\text{crit}}$. In the fundamental diagram, this critical speed separates the free-flow region from the congestion region.
%The free-flow set of location $i$, i.e. the set containing all data points corresponding to free flow, is defined as
%\begin{equation}
    %\mathcal{F}^{(i)}:=\bigcup_{j \in \mathcal{J}^{(i)}} %\mathcal{F}^{(i,j)},
%\end{equation}
%where 
The free-flow set of location $i$ on date $j$, i.e. the set containing all data points corresponding to free flow, is defined as
\begin{equation}
    \mathcal{F}^{(i,j)}:=\bigg\{ \left(q^{(i,j)}_t, v^{(i,j)}_t, \rho^{(i,j)}_t\right) \in \mathcal{X}^{(i,j)} :  v^{(i,j)}_t \geq  v^{(i)}_{\text{crit}} \bigg\},
\end{equation}
i.e. the set of all data points of location $i$ and date $j$ for which the average speed is equal to or higher than the critical speed of location $i$.
Naturally, the congestion set %of location $i$ 
is defined as the complement of the free-flow set, i.e.
%\begin{equation}
%    \mathcal{C}^{(i)}:=\bigcup_{j \in  \mathcal{J}^{(i)}} \mathcal{C}^{(i,j)},
%\end{equation}
%where 
\begin{equation}
    \mathcal{C}^{(i,j)}:=\mathcal{X}^{(i,j)} \setminus \mathcal{F}^{(i,j)}.
\end{equation}

\begin{figure}[!htb]%
 \centering
 \subfloat[]{
     \begin{overpic}[scale=0.55]{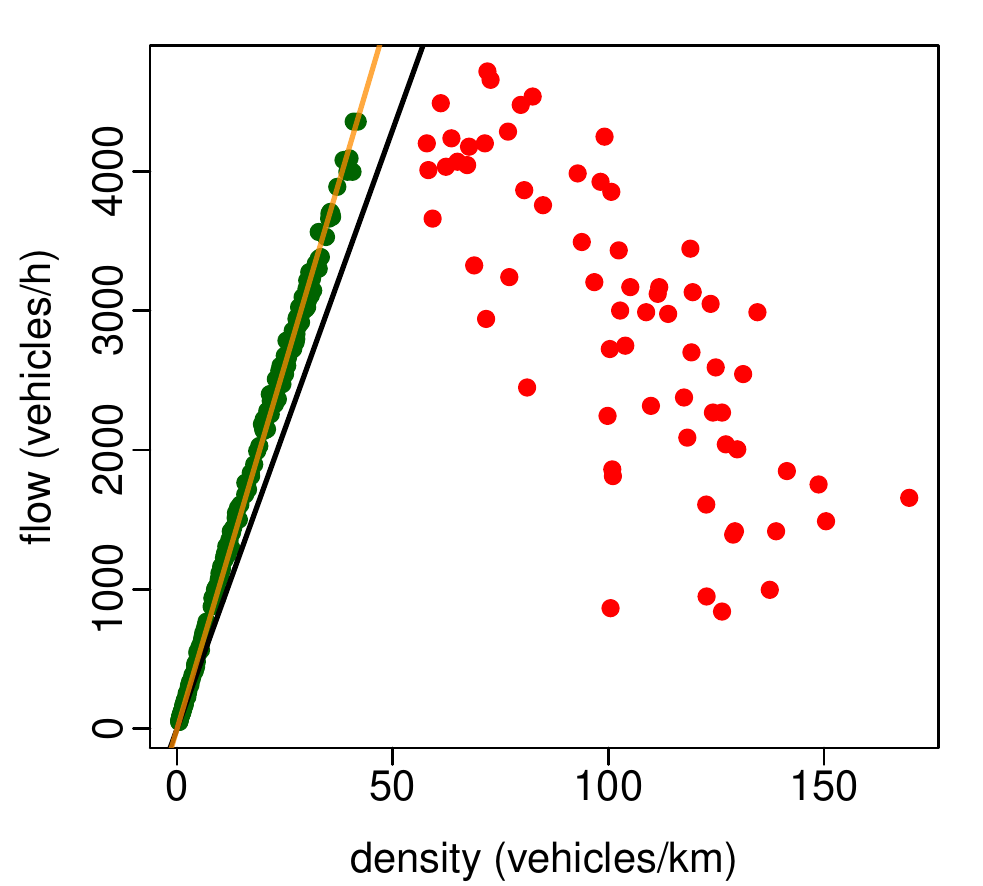}
        \definecolor{darkgreen}{rgb}{0, 0.39, 0}
        \newcommand{\bignwarrow}{\text{\scalebox{2}{\ensuremath{\nwarrow}}}}
         \put(40,40){\textcolor{red}{$\mathcal{C}^{(i,j)}$}}
         \put(18,72){\textcolor{darkgreen}{$\mathcal{F}^{(i,j)}$}}
         \put(25,25){$\bignwarrow$}
         \put(40,20){$v^{(i)}_{\text{crit}}=86$ km/h}
    \end{overpic}\label{fig:2a}}
 \quad
 \subfloat[]{
    \begin{overpic}[scale=0.55]{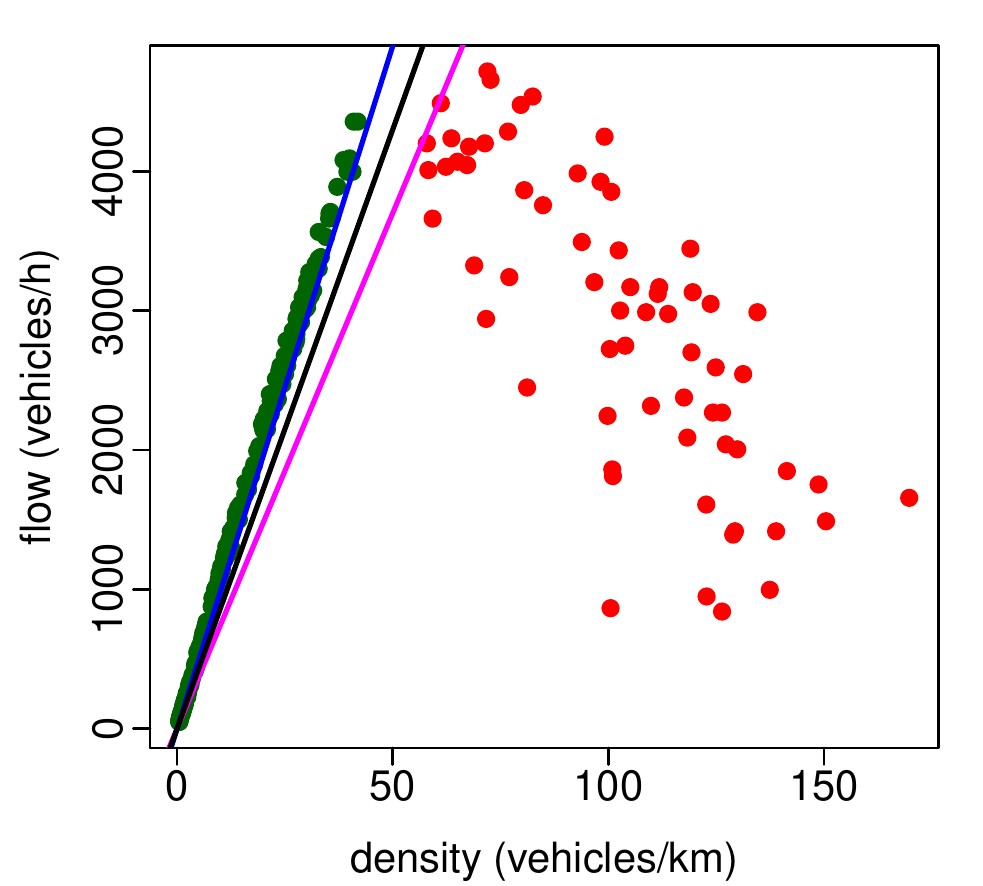}
    \end{overpic}\label{fig:2b}}\\
 \caption{The fundamental diagram with free-flow points (green) and congestion points (red). In (a) it is shown how the free-flow region and the congestion region are linearly separable by a straight line through the origin (the black line), the slope of this line is the critical speed. Additionally, the slope of the orange line through the origin is the (constant) free-flow speed, which is 95.5 km/h. Note that the free-flow speed is significantly below the speed limit, as this is an average over both multiple vehicles and multiple lanes. In (b) it is shown how the critical speed can be estimated by the line that lies exactly between the boundary line of the free-flow region (blue) and the boundary line of the congestion region (magenta).}%
 \label{fig:fund}%
\end{figure}

The difference between free flow and congestion is clearly visible in the fundamental diagram (or the empirical fundamental diagram of traffic flow, to be precise), which is a plot of the measured flow rates $q_t^{(i,j)}$ against the vehicle densities $\rho_t^{(i,j)}$.
An example of the empirical fundamental diagram is presented in Figure~\ref{fig:fund}a. In this example, the black line clearly separates the free flow set from the congestion set.

During free flow, the flow-density relationship can be modelled by a straight line (see the orange line in Figure~\ref{fig:fund}a), which logically must pass through the origin:
\begin{equation}\label{eq:assumptionFreeFlow}
    q \approx \rho\cdot v^{(i)}_{\text{free}} \quad \forall(q,v,\rho) \in  \mathcal{F}^{(i,j)}.
\end{equation}
%Although this is not necessarily the case at our location (see Section~\ref{sec:data}), the data indicates that in our case study this is a reasonable assumption.
When using the data set $\mathcal{X}^{(i,j)}$, we assume the following conditions are met:
\begin{enumerate}[(i)]
    \item The average speed during free flow $v^{(i)}_{\text{free}}$ is constant for all locations $i\in \mathcal{I}$;
    \item The road conditions at location $i$ are homogeneous for all dates $j \in \mathcal{J}^{(i)}$, for all locations $i\in \mathcal{I}$;
    \item For each $i \in \mathcal{I}$ and $j \in \mathcal{J}^{(i)}$, the number of free-flow measurements significantly exceeds the number of congestion measurements.
\end{enumerate}
Whenever at least one of these conditions is violated, for a certain day $j$ at location $i$, day $j$ will not be taken into account when determining $v^{(i)}_{\text{crit}}$. The first condition is rarely violated, since a constant free flow speed follows from the definition of free flow (see e.g. \cite{kerner2009IntroModern}), given conflict free roads with a fixed speed limit and homogeneous conditions. Assumptions (ii) and (iii) may be violated on days where circumstances are completely different from ordinary days, for example in case of accidents, road works or extreme weather conditions. These days could be detected using additional data and therefore be removed from the data set. However, in order to keep the algorithm as simple and self-contained as possible, we simply choose to exclude the most extreme critical speeds. We emphasise that in our experimental region the core elements of the road were fixed throughout the year, i.e. the speed limit is fixed and no traffic lanes where removed or added. Furthermore, despite the experimental region being subject to heavy congestion, congestion occurs mainly during the morning and afternoon rush hour, which means that in general the number of free-flow measurement well exceeds the number of congestion measurements. As a result, Assumptions (i), (ii) and (iii) are only violated in extreme cases and removing the most extreme critical speeds will be sufficient to ensure the assumptions are met. This explains the first point regarding the removal of several critical speeds stated in Section~\ref{sec:data}.

\vspace{-1 cm}

\subsection{Using Robust Regression to Label Data Points}\label{sec:robust}

The purpose of our algorithm is to find the free flow set and the congestion set, for every day and location. More formally, we aim to find a label for each $(q,v,\rho)\in \mathcal{X}^{(i,j)}$ that indicates whether $(q,v,\rho)\in \mathcal{F}^{(i,j)}$ or $(q,v,\rho)\in \mathcal{C}^{(i,j)}$. A logical first step is to determine the straight line through the origin that lies exactly between the free-flow region and the congestion region, as depicted by the black line in Figure~\ref{fig:fund}b. 
The slope of this line is the estimate of the critical speed of location $i$ for each date $j \in \mathcal{J}^{(i)}$, denoted by $v^{(i,j)}_{\text{crit}}$.

In order to obtain the critical speed and the corresponding labelling from the fundamental diagram, several methods have been studied in the literature. Examples are an iterative regression method after performing a change-point analysis~\cite{arnesen2017estimator}, the use of fuzzy logic for clustering~\cite{stutz2002classification}, and assuming a specific model for the fundamental diagram, obtaining the critical density and subsequently labelling each point
~\cite{knoop2017automatic}. However, we opt for a more intuitive and efficient method based on robust regression, to exploit the underlying structure of the fundamental diagram.

Robust regression essentially does the same as ordinary regression, yet is more robust to potential violations of the modelling assumptions (e.g. outliers), see for example~\cite{montgomery2012introduction}.
To this end, each data point $\mathbf{x}$ is assigned a weight $w(\mathbf{x})\in \lbrack 0, 1\rbrack$ and subsequently a linear model is fitted and a reiterative weighted least squares fit is performed (where the weights are updated each step according to the new estimate); in this way outliers have a smaller influence on the final estimates due to their lower weights and the model aims to fit the majority of the data, rather than the whole data set. 
We apply robust regression to the flow-density set $\bar{\mathcal{X}}^{(i,j)}$ of each location $i$ and date $j$ separately. Specifically, we fit the following model:
\begin{equation}\label{eq:regression_model}
    q_t = v^{(i,j)}_{\text{free}}\cdot \rho_t  +\varepsilon_t \quad \forall(\rho_t,q_t) \in  \bar{\mathcal{X}}^{(i,j)},
\end{equation}
where the $\varepsilon_t$ are error terms with expectation zero. %With robust regression, outliers get less weight in the parameter estimation. 
In our case, the ``outliers'' are the points corresponding to congestion.
The reason why this method works so well for this application, is threefold:
\begin{enumerate}
\item We exploit the fact that in free flow, the relation between $q$ and $\rho$ is linear;
\item We do not have to assume any specific relation between $q$ and $\rho$ in the congested set, because these points fulfil the role of outliers;
\item The method computes weights that are a measure for the contribution of each point to the final estimate, which can be used for the labelling.
\end{enumerate}

\begin{remark}\label{rem:n1vsn2}
Assumption (iii) from Subsection~\ref{sec:critical_speed}, specifying that we only consider days where the number of points corresponding to congestion is smaller than the number of free flow points, is essential. On a day where this assumption is violated, we have more points belonging to congestion, meaning that the fitted regression line would no longer pass through the free flow set. In this case, the estimated free flow speed $v^{(i,j)}_{\text{free}}$ would be significantly lower than the maximum speed, which makes these days extremely easy to detect (and remove).
\end{remark}

The robust regression is performed using the function \textsf{rlm} from the \textsf{MASS}-package in R, with MM-estimation and Tukey's Bisquare function for the weights with the default S-estimator as suggested in \cite{venables2013modern}. Tukey's Bisquare function behaves similarly to the squared error function except for larger errors, for which it decreases the weight (see e.g. \cite{montgomery2012introduction}).
This results in an estimate for $v_{\text{free}}^{(i,j)}$ and certain weights $w(\mathbf{x})$ for each data point $\mathbf{x}\in \bar{\mathcal{X}}^{(i,j)}$. Instead of the usual interest in the model and parameter estimation, we are interested in the \emph{weights} associated with each data point. Using the weights, we perform the labelling: if the weight is low and if the data point corresponds to a speed lower than the free-flow speed, $v_{\text{free}}^{(i,j)}$, the data point will be labelled as congestion. All other points will be labelled as free flow. 
Hence, for each $\mathbf{x}=\left(\rho,q\right)\in \bar{\mathcal{X}}^{(i,j)}$ we determine $\mathds{1}_\mathcal{C}(\mathbf{x}):=\mathds{1}\lbrace\mathbf{x}\equiv \left(q,v,\rho \right)\in \mathcal{C}^{(i,j)}; \mathbf{x}\in \bar{\mathcal{X}}^{(i,j)} \rbrace$, i.e. the indicator function for the event that $\mathbf{x}$ corresponds to congestion or not. The critical weight has been placed at $0.01$ (see Section \ref{sec:validation} for a justification), hence
\begin{equation}\label{eq:indicatorCongestion}
      \mathds{1}_\mathcal{C}(\mathbf{x})= 
\begin{cases}
    1& \text{if } w(\mathbf{x})< 0.01 \text{ and } v=q/\rho < v_{\text{free}}^{(i,j)}\\
    0              & \text{otherwise}.
\end{cases}
\end{equation}

After we obtain the labels, we estimate $v_{\text{crit}}^{(i,j)}$ (see the black line in Figure~\ref{fig:fund}b) by determining the slope of the straight line through the origin that lies exactly between the free-flow region and the congestion region.
\begin{remark}\label{rem:not_separable}
It may happen that the boundary line of the congestion region lies above the boundary line of the free-flow region (i.e. the magenta line has a larger slope than the blue line in Figure~\ref{fig:fund}b), since the weights are calculated based on the Euclidean distance from the free-flow line. In this case, the free-flow region and the congestion region are not linearly separable by a straight line through the origin, given the labelling. For such instances, there will exist data points $\mathbf{x}\equiv\left(q,v,\rho \right) \in \mathcal{C}^{(i,j)}$ and $\mathbf{x'}\equiv\left(q',v',\rho' \right) \in \mathcal{F}^{(i,j)}$ such that $v>v'$. The critical speed for such instances is indeterminate and therefore we do not include these instances in the determination of the critical speed of the corresponding location.
\end{remark}
In the end, the critical speed of location $i$ is estimated as follows:
\begin{equation}\label{eq:critSpeedMedian}
    v_{\text{crit}}^{(i)}=\text{median}\lbrace \mathcal{V}_{\text{crit}}^{(i)} \rbrace,
\end{equation}
where
%\begin{equation}\label{eq:sigmaDistance}
%    \mathcal{V}_{\text{crit}}^{(i)}:=  \bigg\{  v_{\text{crit}}^{(i,j)} : \left\vert v_{\text{crit}}^{(i,j)} - %\mu\big\{v_{\text{crit}}^{(i,j)}\big\}_{j\in \mathcal{J}^{(i)}} \right\vert<2\cdot \sigma\big\{v_{\text{crit}}^{(i,j)}\big\}_{j\in %\mathcal{J}^{(i)}}  \bigg\},
%\end{equation}
\begin{align}
        &\mathcal{V}_{\text{crit}}^{(i)}:=  \bigg\{v_{\text{crit}}^{(i,j)} \bigg\}\label{eq:set_crit}&&\\
        \text{such that:}\nonumber  \\
        &\left\vert v_{\text{crit}}^{(i,j)} - \mu\big\{v_{\text{crit}}^{(i,j)}\big\}_{j\in \mathcal{J}^{(i)}} \right\vert<2\sigma\big\{v_{\text{crit}}^{(i,j)}\big\}_{j\in \mathcal{J}^{(i)}};\label{eq:2sigma}\\
        &v'>v\quad \forall \mathbf{x} \in \mathcal{C}^{(i,j)}, \mathbf{x'} \in \mathcal{F}^{(i,j)};\label{eq:boundary_wrong}\\
        &\text{MAPE}\left(\bar{\mathcal{X}}^{(i,j)}\right)\geq 0.1 \label{eq:no_congestion}.
\end{align}
where $\mu \lbrace \cdot \rbrace$ and $\sigma \lbrace \cdot \rbrace$ denote the mean and standard deviation of the corresponding sets respectively and $\text{MAPE}\left(\bar{\mathcal{X}}^{(i,j)}\right)$ denotes the mean absolute percentage error of the regression model presented in Equation~\eqref{eq:regression_model}.

Equation~\eqref{eq:2sigma} removes the most extreme critical speeds. By excluding days with a critical speed that lies outside a range of twice the standard deviation from the average, we prevent potential violations of the assumptions from influencing the estimates (as elaborated upon in Section~\ref{sec:critical_speed}). Equation~\eqref{eq:boundary_wrong} excludes days where the boundary line of the congestion region lies above the boundary line of the free-flow region (see Remark~\ref{rem:not_separable}). Lastly, Equation~\eqref{eq:no_congestion} ensures that the critical speed of a location is not based on days with little or no congestion. As one can imagine, in case of hardly any congestion, a free-flow point with a relatively slow speed might be incorrectly labelled as congestion. We therefore impose a minimal level of congestion and use the mean absolute percentage error (MAPE, see e.g. \cite{MAPE}) of the corresponding model (see Equation~\eqref{eq:regression_model}) as a surrogate of the average congestion level. The MAPE expresses the error of the model in terms of a percentage; a low MAPE corresponds to a very accurate model, implying hardly any congestion, whereas a high MAPE indicates that various points deviate from the straight line through the origin, which corresponds to the presence of congestion during that day. The critical level of the MAPE has been placed at 0.1, in Section~\ref{sec:validation} this threshold will be motivated.

The set of critical speeds of location $i$, corresponding to the instances of location $i$ which satisfy the three conditions presented in Equations~\eqref{eq:2sigma},~\eqref{eq:boundary_wrong} and~\eqref{eq:no_congestion}, is given by $\mathcal{V}_{\text{crit}}^{(i)}$. The critical speed of location $i$ is subsequently determined by taking the median of this set. We take the \emph{median} of the critical speeds among multiple days to provide a solid baseline for comparison among different days. We emphasise that in the end the critical speed of each location is estimated as the median of at least 147 critical speeds (out of 261 weekdays) and that most instances were removed based on Equation \eqref{eq:boundary_wrong}.

%two other cases. It may happen that the boundary line of the congestion region lies above the boundary line of the free-flow region (i.e. the magenta line has a higher slope than the blue line in Figure~\ref{fig:fund}b), since the weights are calculated based on the Euclidean distance from the free-flow line. In such cases, the critical speed of that day will be discarded immediately. This reduces the total number of instances used in the first step to 871. Finally, removing the most extreme critical speeds of each location, according to Equation~\eqref{eq:sigmaDistance}, brings the total number of utilized critical speeds to 839 (roughly equally distributed over the five locations). In the end, the critical speed of each location is estimated as the median of at least 145 critical speeds.

\subsection{Estimating the Breakdown Probability and Identifying the High-performance Days} \label{sec:breakdown_high-performance}

Congestion arises as a consequence of a breakdown, which is defined as a transition from free flow to congestion (see, e.g.~\cite{arnesen2017estimator}). Usually, this happens when the traffic flow is high and some kind of disruption occurs (e.g. a vehicle changing lanes or another sudden movement of a driver). 

\begin{definition}[Breakdown]\label{def:breakdown}
A breakdown, at location $i$ and date $j$, is a moment $t^{(i,j)}_k \in \mathcal{T}^{(i,j)}$ such that
\begin{equation*}
    v^{(i,j)}_{t^{(i,j)}_k}\geq v_{\text{crit}}^{(i)}>v^{(i,j)}_{t^{(i,j)}_{k+1}}.
\end{equation*}
\end{definition}

We assume that breakdowns have a probabilistic nature, see e.g.~\cite{arnesen2017estimator,tu2008monitoring}, meaning that from a macroscopic point of view the occurrence of breakdowns (given a certain traffic flow) is random. This implies the existence of a breakdown probability (as a function of the traffic flow). To estimate this probability, we use the non-parametric estimator discussed in Arnesen and Hjelkrem~\cite{arnesen2017estimator}. To calibrate this estimator,  the aforementioned classification of each data point as either free flow or congestion is required. Arnesen and Hjelkrem define two functions; $Q^{(i)}(q)$, which is the number of breakdowns at location $i$ while the traffic flow is equal to or lower than $q$, and $R^{(i)}(q)$, which is the number of times a breakdown did not occur at location $i$ with a traffic flow of at least $q$. Subsequently, the breakdown probability $P^{(i)}(q)$, which denotes the probability of a breakdown at location $i$ when the traffic flow is $q$, can be estimated by
\begin{equation} \label{eq:ecdf_breakdown}
    P^{(i)}(q) = \frac{Q^{(i)}(q)}{Q^{(i)}(q)+R^{(i)}(q)}.
\end{equation}

\begin{remark}\label{rem:fake_breakdowns}
To avoid including ``fake breakdowns'' (e.g. a single vehicle driving unnecessarily slow at night), we pose the additional constraint on a breakdown that it does not happen before 5:00 in the morning. Indeed, multiple times we observed before 5:00, at a minimal traffic flow, a sudden drop of the average speed to just below the critical speed. We assume that such events are not relevant for estimating the breakdown distribution as this could be a truck driving at its speed limit of 80 km/h.
\end{remark}

To reduce the complexity of the estimation method, we use a surrogate for the breakdown probabilities, obtained by fitting a cumulative normal distribution function, as is done in~\cite{arnesen2017estimator}.

In Section~\ref{sec:intro}, an intuitive description of a high-performance day was given. In this section we present a criterion to determine a quantitative definition for high-performance days. To this end, we employ the estimated breakdown probability in Equation~\eqref{eq:ecdf_breakdown}, to find \emph{unperturbed moments}. An unperturbed moment is a moment at which the probability of a breakdown is at least $0.5$, but the expected breakdown did not occur, or more mathematically:

\begin{definition}[Unperturbed moment]
An unperturbed moment, at location $i$ on date $j$, is a moment $ t^{(i,j)}_k \in \mathcal{T}^{(i,j)}$  with intensity $q^{(i,j)}_{t^{(i,j)}_k}\geq q_{\text{upt}}^{(i)}$ and speed $v^{(i,j)}_{t^{(i,j)}_k}\geq v^{(i)}_{\text{crit}}$ for which it holds that
\begin{equation}
P^{(i)}\big(q^{(i,j)}_{t_k^{(i,j)}}\big) \geq 1/2 \quad  \wedge \quad v^{(i,j)}_{t^{(i,j)}_{k+1}}\geq v^{(i)}_{\text{crit}},
\end{equation}
where $q_{\text{upt}}^{(i)}$ is the smallest value of the traffic flow $q$ such that $P^{(i)}(q)\geq 1/2$.
\end{definition}
A plausible definition of a high-performance day follows naturally.
\begin{definition}[High-performance day]
A high-performance day is a day with a large number of consecutive unperturbed moments in both time and space compared to other days.
\end{definition}

Note that a high-performance day is thereby a relative measure, as it will depend on the location how many unperturbed moments are generally present (some locations experience more variability in terms of breakdowns in relation to the traffic flow). Indeed, a certain level of freedom in the definition of high-performance days is required. For example, quantifications such as the top 0.05 percentile, though plausible in some cases, incorrectly imply the existence of high-performance days at any location. Furthermore, concretizations of the definition in terms of the number of unperturbed moments depend on the experimental region.

\section{Key Insights and Validation} \label{sec:results}
In this section, we present the results of our algorithm and validate the estimation methods. In particular, we study the results of the three steps of the algorithm and present several measures of the top 10 high-performance days. In addition, we take a closer look at what exactly a high-performance day looks like and how we can use our macroscopic data to visualise the dynamics of such days for the whole trajectory. Subsequently, we elaborate on several problems one might encounter when applying the method at a different location and how these problems could be tackled. Specifically, we state how we dealt with these problems and how we obtained the critical weight and the critical level of the MAPE.

\subsection{Results and Key Insights} \label{sec:insights}

In Table~\ref{tab:results_alg} we present the results of the first and second step of the algorithm (i.e. estimating the critical speed and the breakdown probabilities respectively). We observe that the critical speed is roughly equal for the various locations. We see a similar behavior for the estimated free-flow speeds, which are consistently roughly 10 km/h above the corresponding estimated critical speeds. Furthermore, we observe that the smallest value of the traffic flow for which the breakdown probability is at least 0.5 decreases along the trajectory, meaning that the last two locations experience breakdowns at a lower traffic flow than the first three locations. This makes sense considering the merge from 3 lanes to 2 lanes at the fourth location.

\begin{table}[ht!]
\vspace{-0.5cm}
\caption{Columns from left-to-right: the rounded estimated critical speed of location $i$, the rounded estimated free-flow speed of location $i$ (based on the median of the free-flow speeds of the instances that were used to estimate the critical speed of location $i$), the number of instances used for estimating the critical speed of location $i$ (out of a total of 261 weekdays) and the smallest traffic flow for which the breakdown probability is at least 0.5. The speeds are expressed in kilometres per hour and the traffic flows are expressed in vehicles per hour.}
\centering
\begin{tabular}{C{2.2cm}C{1.0cm}C{1.0cm}C{1.0cm}C{1.0cm}}
  & $v^{(i)}_{\text{crit}}$& $v^{(i)}_{\text{free}} $& $\vert \mathcal{V}^{(i)}_{\text{crit}}\vert$& $q^{(i)}_{\text{upt}}$ \\
\hline
Location 1 & 95.5  & 104.5   & 147& 4358\\ 
Location 2 & 93    & 103   & 162& 4019\\ 
Location 3 & 93    & 102   & 175& 3901\\ 
Location 4 & 94.5 & 104.5   & 180& 3195\\ 
Location 5 & 92.5 & 102.5  &175& 3164\\
\end{tabular}
\label{tab:results_alg}
\end{table}

\begin{figure}[ht!]
    \centering
    \includegraphics[width =0.9 \linewidth]{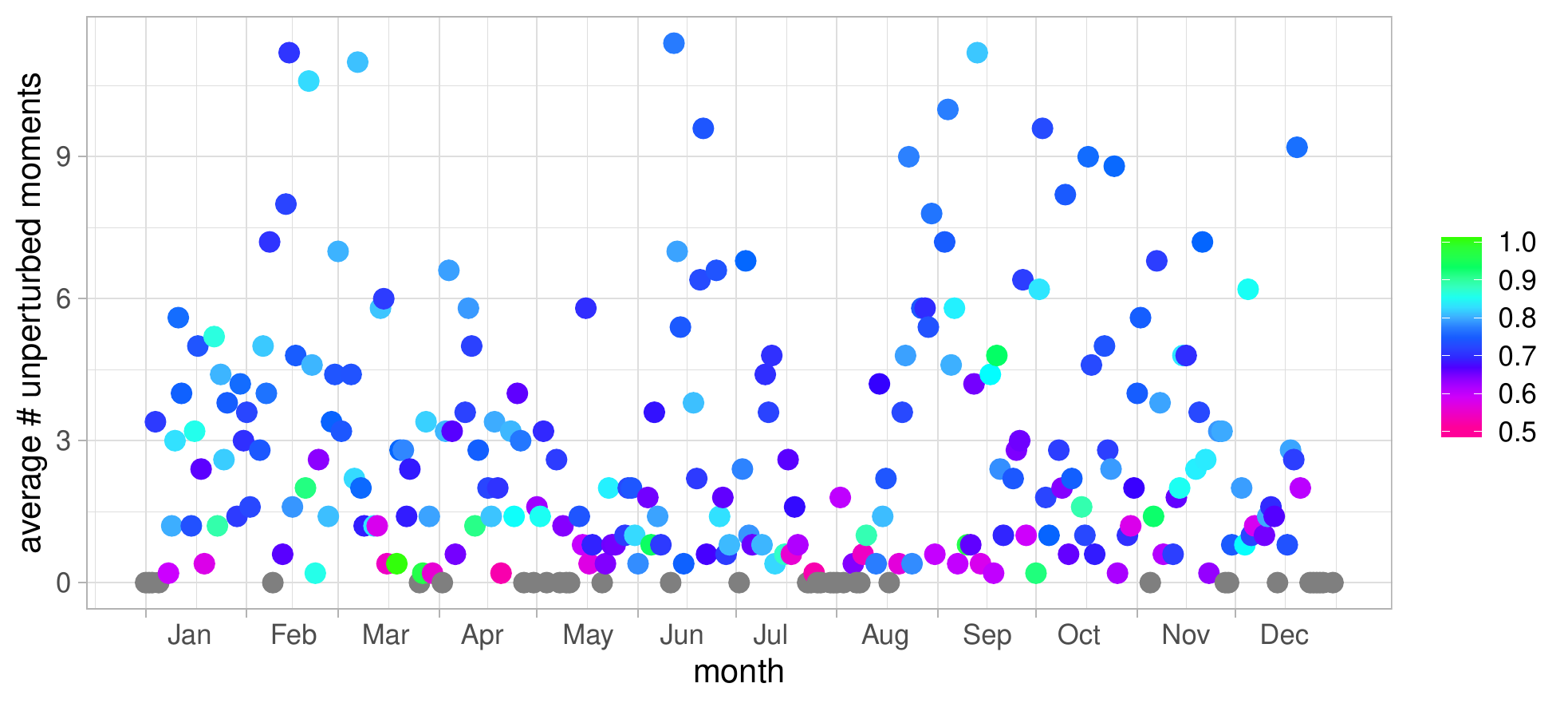}
    \caption{Plot of the average number of unperturbed moments for each weekday of 2018. The colour of each point indicates the average breakdown probability of the unperturbed moments. In case no unperturbed moments occurred, the corresponding point is grey.}
    \label{fig:unperturbedMomentsTimeSeries}
\end{figure}

In Figure~\ref{fig:unperturbedMomentsTimeSeries} we present a scatter plot displaying the average number of unperturbed moments per location for each weekday of 2018. Additionally, the colour of each point corresponds to the average breakdown probability of the unperturbed moments. We observe that the days can be grouped into roughly three categories: days with hardly any unperturbed moments, days with some unperturbed moments, and days with a relatively large number of unperturbed moments. %Of course, where one draws the boundary line between these three groups is subjective. 
It turns out that most days in the first group correspond to days with significantly less traffic, thus implying a low traffic flow and thereby a lack of unperturbed moments. For example, the grey points in Figure~\ref{fig:unperturbedMomentsTimeSeries} often correspond to (school) holidays. The third group, however, is of major interest to us, as these are the high-performance days.

In Table~\ref{tab:table} we present several measures of the top 10 high-performance days (based on Figure~\ref{fig:unperturbedMomentsTimeSeries}), corresponding to the fourth location. We choose to only present results for the fourth location, because averaging the speeds over the various locations requires a critical speed for the whole trajectory as a baseline (whose definition is not straightforward). To study the characteristics of these days, we investigate the average speed and average fraction of free-flow measurements. We look at three time intervals: the morning rush hour 6.30-9.30, outside peak hours 9.30-15.30 and the afternoon rush hour 15.30-19.00. We observe that, though all days show a relatively large number of unperturbed moments, the characteristics of the various days can differ greatly. For example, the top 7 high-performance days all have an average speed during the morning rush hour that is below the critical speed of the corresponding location (i.e. 94.5 km/h) and at least 10\% of the measurements during the morning rush hour correspond to congestion, whereas the remaining three days shows\ hardly any signs of congestion in the morning. We also observe a similar pattern across all high-performance days; the mornings are significantly better (in terms of the average speed and the fraction free flow) than the afternoons. In fact, it seems that severe congestion during the afternoon was present in almost all high-performance days (only February~14, 2018 is an exception). Nevertheless, the mornings of the top 10 high-performance days are quite extraordinary, in particular when comparing the average speed and the fraction free flow with the median over all weekdays from 2018 at location 4.

\begin{table}[ht!]
    \vspace{-0.5cm}
    \caption{Several measures of the top 10 high-performance days, based on Figure~\ref{fig:unperturbedMomentsTimeSeries}, corresponding to the fourth location. The average speed is presented during the morning rush hour 6.30-9.30, outside peak hours 9.30-15.30 and during the afternoon rush hour 15.30-19.00, as well as the corresponding fraction free flow. The median over all weekdays of 2018 is presented as well.\\}
    \centering
    \includegraphics[width =1 \linewidth]{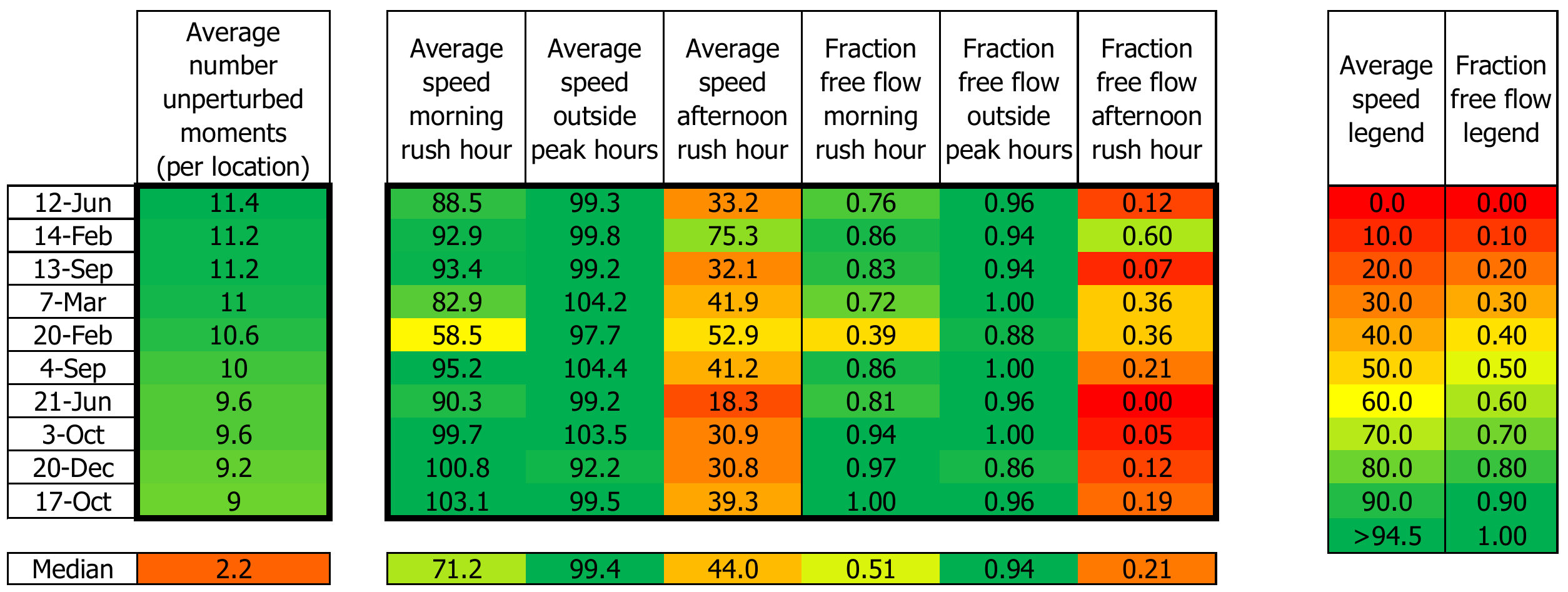}
    \label{tab:table}
\end{table}

\begin{figure}[ht!]
    \centering
    \includegraphics[width = 0.9\linewidth]{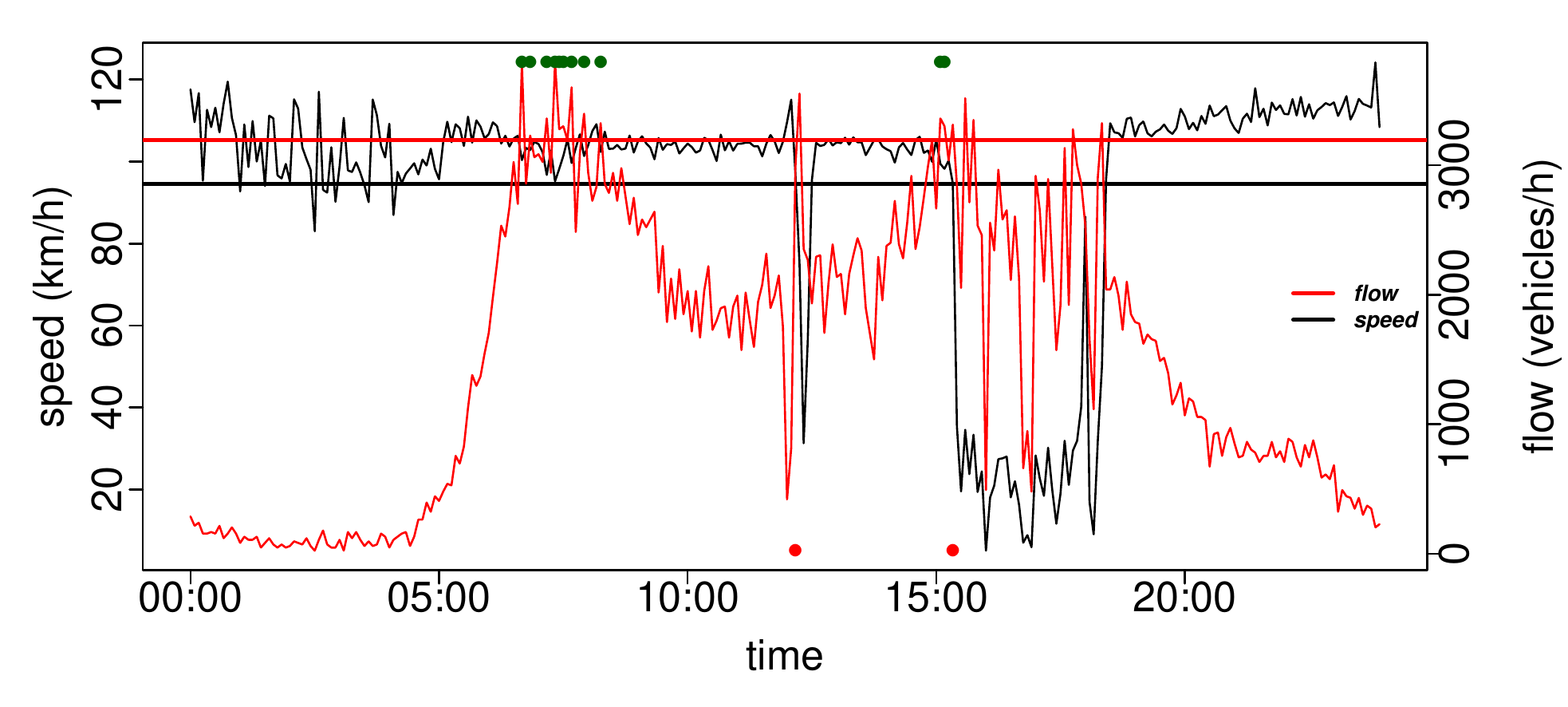}
    \caption{Time series of the average speed (black) and average flow (red) during October~17, 2018 at location 4. Unperturbed moments are indicated by a green dot and breakdowns are indicated by a red dot. The horizontal black line is the estimated critical speed and the horizontal red line is the smallest traffic flow for which the breakdown probability is at least 0.5.\\}
    \label{fig:timeSeriesSpeedFlow}
\end{figure}

We now thoroughly study the traffic behaviour during October~17, 2018. During this day, an average of 9 unperturbed moments was identified (see Table~\ref{tab:table}). This day is particularly interesting because of the seemingly large difference between the morning rush hour and the afternoon rush hour. In fact, this day is the only day in the top 10 high-performance days which does not show any congestion during the entire morning rush hour. In Figure~\ref{fig:timeSeriesSpeedFlow} a joint time series of the average flow and average speed at the fourth location during this day is presented. As expected, we notice a large number of unperturbed moments, mostly during the morning. The contrast between the morning and the afternoon is indeed interesting, as the breakdown, which remained absent in the morning, manifested in the late afternoon at a lower traffic flow. This is in line with our probabilistic view on the occurrence of a breakdown, at least from a macroscopic point of view, and confirms that this morning was indeed extraordinary.

\begin{figure}[ht!]%
 \centering
 \includegraphics[width = 0.9\linewidth]{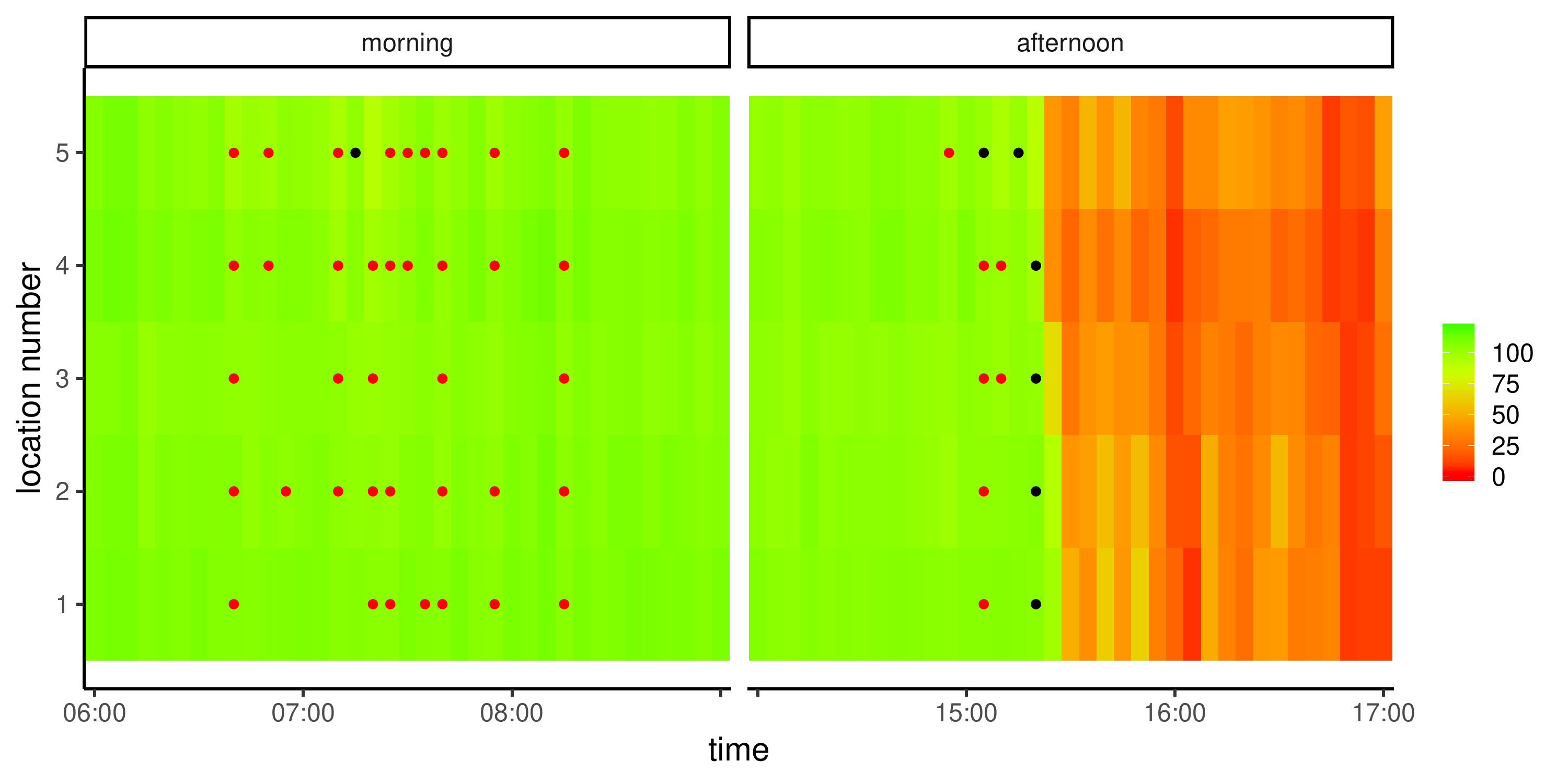}
 \caption{A space-time diagram of the morning rush hour and the afternoon of October 17, 2018. The average speed is displayed along the whole trajectory. Furthermore, breakdowns are marked with a black marker and unperturbed moments are marked with a red dot.}%
 \label{fig:space_time_breakdown1}%
\end{figure}

Additionally, one could employ visualizations to investigate the whole trajectory simultaneously, see Figure~\ref{fig:space_time_breakdown1}. We verified that the morning of October~17, 2018 was extraordinary at the fourth location and Figure~\ref{fig:space_time_breakdown1} shows that this was the case for the whole trajectory. Indeed, we observe multiple unperturbed moments during the morning rush hour at each of the five locations. In particular, despite the high traffic flow (recall that unperturbed moments only occur at a traffic flow of at least 3164 vehicles per hour, see Table~\ref{tab:results_alg}), we observe no significant speed decrease. Furthermore, as we expect based on Figure~\ref{fig:timeSeriesSpeedFlow}, a breakdown along the whole trajectory can clearly be seen around 15.20-15.30 (see Figure~\ref{fig:space_time_breakdown1}).

\subsection{Validation} \label{sec:validation}

The critical speeds are estimated based on a labelling of the data points resulting from the robust regression method discussed in Section~\ref{sec:robust}. As the exact shape of the fundamental diagram depends on the location, it is difficult to make general statements about the accuracy of the critical speed estimation. However, we can identify three possible issues: 1. little or no congestion occurred during a day; 2. extreme congestion occurred during a day; 3. the free-flow speed was not (approximately) constant. We also present a way to determine whether or not those problems did arise (besides additional information about the experimental region). Finally, we conclude this section with a discussion on how to choose the critical weight, which is used to determine whether observations belong to the congestion set or the free-flow set.

\begin{figure}[ht!]
\centering
\begin{minipage}[b]{0.45\linewidth}
\includegraphics[width=\linewidth]{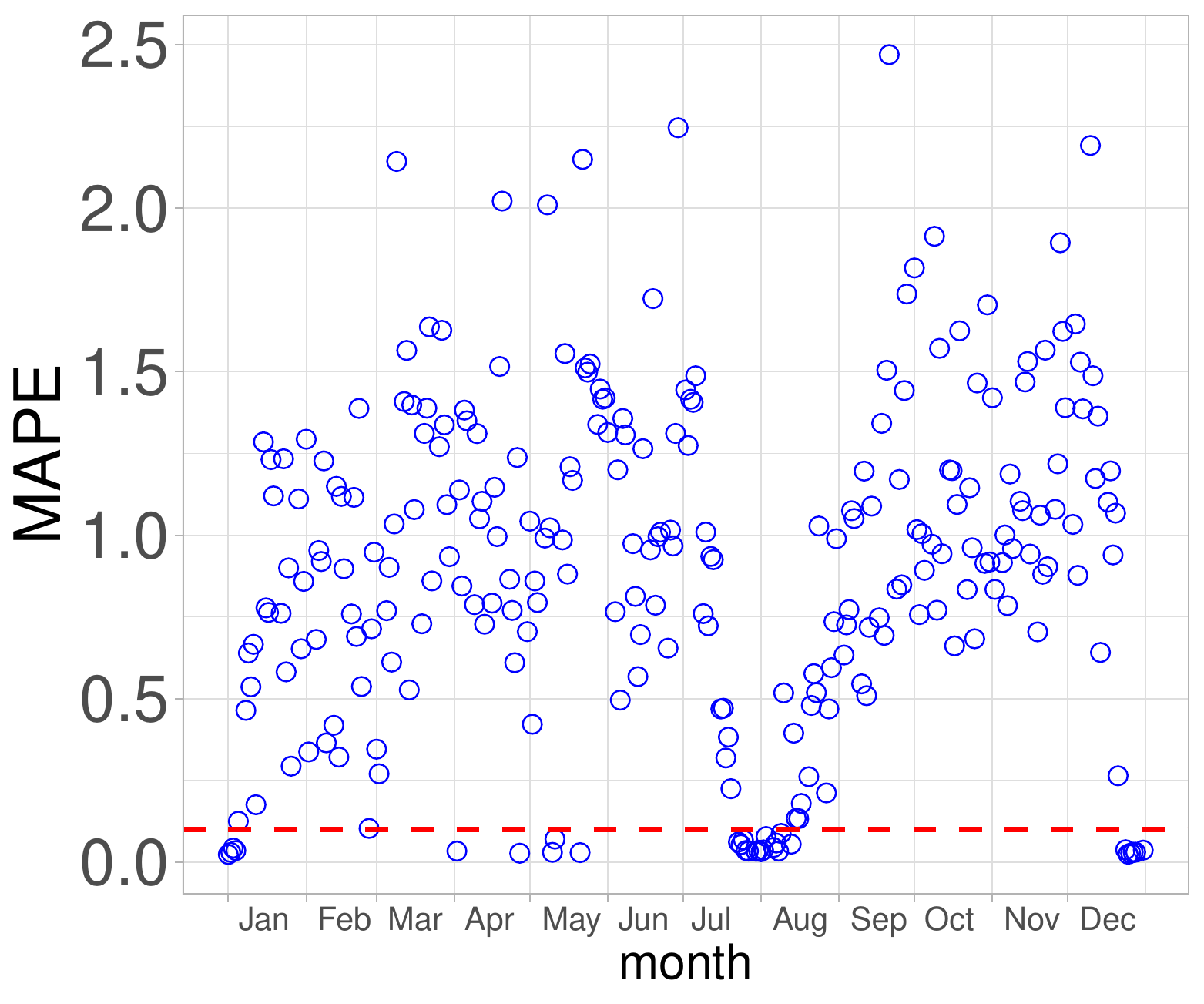}
\caption{Plot of the MAPE of the robust regression model for all days.}
\label{fig:scatterplotError}
\end{minipage}
\hspace{.5cm}
\begin{minipage}[b]{0.45\linewidth}
\centering
\includegraphics[width=\linewidth]{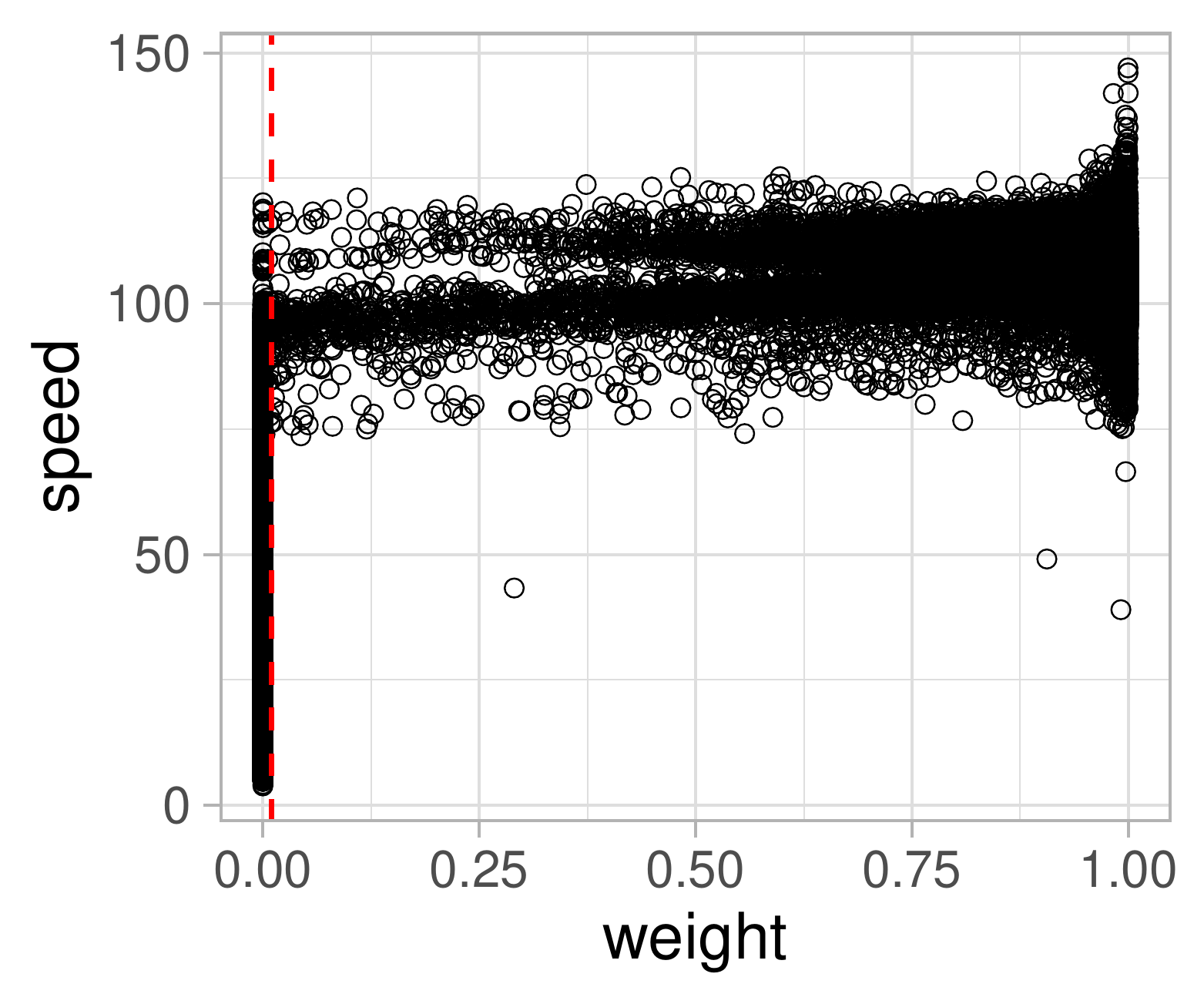}
\caption{Plot of the weights and corresponding speeds for location 1.}
\label{fig:scatterplotWeightSpeed}
\end{minipage}
\end{figure}

\paragraph{Little or no congestion.} In this case, robust regression might interpret a free-flow point with a relatively slow speed as an outlier and therefore cause a free-flow point to be labelled as a congestion point. This leads to a higher estimate of the critical speed during that day. Though in our case it is not likely that the final estimate of the critical speed will be strongly influenced by several overestimates (considering that our experimental region is generally subject to heavy congestion), we still exclude days with little or no congestion. As mentioned in Section~\ref{sec:robust}, we use the mean absolute percentage error (MAPE) of the robust regression model presented in Equation~\eqref{eq:regression_model} as a surrogate for the average congestion level. In Figure~\ref{fig:scatterplotError} a scatter plot of the MAPE for the various days of location 1 is shown. We observe that, for example, during the holidays (the beginning of January/end of December) and throughout the summer break, the MAPE is close to zero. Indeed, during those days the traffic flow was significantly lower and therefore hardly any congestion occurred. Based on Figure~\ref{fig:scatterplotError} (and similar figures for the other locations), we decided to place the threshold at 0.1; instances with a MAPE of less than 0.1 will be excluded when determining the critical speed, as in Equation~\eqref{eq:no_congestion}.

\paragraph{Extreme congestion.} This may lead to severe underestimations of the critical speed. One can imagine that if the number of congestion measurements becomes too large, not all congestion points will be observed as outliers by the robust regression method. In particular, what may happen is that robust regression fits a model through the congestion region, see also Remark~\ref{rem:n1vsn2}. For the MM-estimators it is known that (asymptotically) in case more than half of the data points lie on a straight line through the origin, the final model will fit that line \cite{yohai1987high}. This means that, if we assume a constant flow-density relation in free flow, the free-flow speed should be accurately estimated if more than half of the measurements correspond to free-flow. However, because Equation~\eqref{eq:assumptionFreeFlow} is only an approximate relation, the algorithm will be even more sensitive to a larger congestion set. In our case study, the fraction of free flow was generally well above 0.5. However, before employing robust regression to determine the critical speed, it is recommended one verifies that the average free-flow level is above 0.5. In case the congestion level is around 0.5 one should cautiously verify that the critical speed is correctly estimated (by e.g. studying the distribution of the estimated critical speed for the various days).

\paragraph{Non-constant free-flow speed.} In case the free-flow speed is not constant, the structure of the fundamental diagram will change heavily (in comparison with e.g. Figure~\ref{fig:fund}). One example would be a decrease of the speed limit when the rush-hour lane is open. In case the rush-hour lane is opened during peak hours, this could result in a free-flow curve, rather than a straight line, displaying an average speed decrease at high traffic flows. Such a scenario could be problematic for our algorithm, as the approximate flow-density relationship, presented in Equation~\eqref{eq:assumptionFreeFlow}, no longer holds. We suggest that one beforehand verifies that the free-flow speed is constant, either by using information about the experimental region or by studying the fundamental diagram. In our case there was no dynamic speed limit and the fundamental diagrams showed no indication of a non-constant free-flow speed.

\paragraph{Critical weights. } In Section \ref{sec:robust}, we introduced the critical weight, which is used to distinguish between congestion and free flow. The critical weight has been placed at $0.01$, meaning that points with a weight below $0.01$ are labelled as free flow. This value is determined using Figure~\ref{fig:scatterplotWeightSpeed}, which shows a scatter plot of all speeds and corresponding weights of the first location. We observe that almost all low speeds (say speeds below 70 km/h), have a weight which is either zero or very close to zero. Speed-weight plots of the other four locations showed a similar pattern. Therefore, we conclude that a critical weight of 0.01 generally allows for a sensible labelling.

\section{Conclusion and Discussion} \label{sec:concl}
%\textcolor{green}{move '(the critical speed can even be determined per lane, but we do not pursue this here)' to section 5}

%\textcolor{green}{For a good day, we require both a little or no traffic jams, as well as a high traffic intensity. Indeed, such days are interesting to study: why did no (or only few) traffic jams occur during that day, while the traffic intensity was high? If one can find the reason(s) why, one can also try to improve traffic flow on other days, which is in the end the reason why this study was set up. }

We have developed an algorithm to identify high-performance days based on an estimation of critical speed and the breakdown probability. The algorithm is relatively straightforward and only requires two quantities; the average traffic flow and the average speed. The algorithm relies on the shape of the fundamental diagram; each observation is classified as either free flow or congestion using robust regression and the critical speed is estimated as the separating line between the two sets. Using a non-parametric estimator for the breakdown probability, we are able to quantify both characteristics of a high-performance day (roughly speaking, high speed and high flow). The algorithm has shown its capabilities by identifying high-performance days on the A15 near Papendrecht in 2018.

A natural follow-up question would be in the direction of causality. Indeed, one could wonder \emph{why} certain days exhibit extraordinary behaviour, in terms of an unexpected absence of traffic jams. A possible explanation could be traffic homogeneity; perhaps during the high-performance days, there were fewer trucks, leading to fewer speed differences between vehicles. Alternatively, the answer may lie hidden in microscopic data; certain (desirable) behavioural characteristics of drivers might be over-represented during high-performance days. This paves the way towards reducing traffic jams from a different perspective and may lead to new insights as well as an easier investigation of countermeasures against traffic jams. This non-trivial extension is, however, beyond the scope of this paper. Instead, we present this tool to facilitate further research into countermeasures against traffic jams, as the algorithm is able to identify which days need to be studied further.

We must be critical of our approach as well, in particular in terms of generality. This mainly relates to the two (subjective) thresholds: the critical weight (to distinguish between congestion measurements and free flow measurements) and the critical level of the MAPE of the regression model (to identify a lack of congestion). Both values were determined based on the five locations of the A15 Papendrecht 2018 data set. However, when testing the algorithm on other data sets, we still observed both a sensible labelling of the data points as well as a plausible recognition of days with little or no congestion. In fact, we tested the algorithm on data sets which violated the assumption of a constant free-flow speed and the algorithm still identified days with a high traffic flow and a striking absence of traffic jams.

\subsubsection{Acknowledgements.} This work was supported by NWO under Grant 438-13-206. We thank De Verkeersonderneming for hosting Bo Klaasse during his internship. We thank Stella Kapodistria and Onno Boxma for interesting discussions on the manuscript.

\bibliographystyle{splncs04}
\bibliography{references}
 
\end{document}